\documentclass[11pt,a4paper]{llncs}
\usepackage{fullpage,graphicx}

\newcommand{\occ}
    {\ensuremath{\mathit{occ}}}
\newcommand{\Oh}[1]
    {\ensuremath{\mathcal{O}\!\left({#1}\right)}}

\begin{document}

\title{AliBI: An Alignment-Based Index for Genomic Datasets}
\author{Hector Ferrada\inst{1}
    \and Travis Gagie\inst{2}\fnmsep\inst{3}
    \and Tommi Hirvola\inst{4}
    \and Simon J. Puglisi\inst{3}}
\institute{Department of Computer Science, University of Chile, Chile
    \and Helsinki Institute for Information Technology, Finland
    \and Department of Computer Science, University of Helsinki, Finland
    \and Department of Computer Science and Engineering, Aalto University, Finland}
\maketitle

\begin{abstract}
With current hardware and software, a standard computer can now hold in RAM an index for approximate pattern matching on about half a dozen human genomes.  Sequencing technologies have improved so quickly, however, that scientists will soon demand indexes for thousands of genomes.  Whereas most researchers who have addressed this problem have proposed completely new kinds of indexes, we recently described a simple technique that scales standard indexes to work on more genomes.  Our main idea was to filter the dataset with LZ77, build a standard index for the filtered file, and then create a hybrid of that standard index and an LZ77-based index.  In this paper we describe how to our technique to use alignments instead of LZ77, in order to simplify and speed up both preprocessing and random access.
\end{abstract}

\section{Introduction} \label{sec:introduction}

High-throughput sequencing methods may soon turn genomics into a victim of its own success.  Sequencing the first human genome took over a decade and cost billions of dollars; sequencing a human genome now takes only a day and costs less than a thousand dollars.  This advance will soon lead to datasets of thousands of genomes: e.g., the UK government has announced plans to sequence a hundred thousand of its citizens' genomes.  With current hardware and software, however, a standard computer can hold in RAM an index for approximate pattern matching on only about half a dozen human genomes.

The current standard indexes for genomics --- e.g., Bowtie~\cite{LTPS09}, BWA~\cite{LD09,LD10} and SOAP3~\cite{Liu+12} --- are based on augmented FM-indexes~\cite{FM05}, which are in turn based on the Burrows-Wheeler Transform (BWT)~\cite{BW94}.  The BWT was originally introduced as a tool for data compression and current indexes use this to reduce their space usage.  BWT-based compression does not take full advantage of repetitive structure~\cite{MNSV10}, however, so these indexes' space usage increases nearly linearly with the number of genomes stored.  In contrast, since human genomes are very similar to each other, the LZ77~\cite{ZL77} and similar compression algorithms can store hundreds or thousands of human genomes in RAM.

Several researchers (e.g.,~\cite{ANS12,CN12,DJSS12,GGKNP12,KN13,MNSV10,MNKS13,NPCHIMP13,YWLWX13}) have investigated LZ- and grammar-based indexes, or other indexes designed to work well on repetitive datasets.  Some of these indexes seem quite promising but almost all --- \cite{YWLWX13} is the only exception we know of --- so far support only exact pattern matching.  Exact pattern matching can be used to implement approximate pattern matching with such indexes~\cite{RNOM09} but this implementation takes time to make practical.  Given that it took several years for the FM-index to be augmented for and adopted by the bioinformatics community, we expect it will take at least another three years before LZ- and grammar-based indexes are in widespread laboratory use.

In the meantime, instead of using completely new indexes, we believe scientists should continue to use standard indexes but combined with techniques that scale them to work on more genomes.  We recently described one such technique~\cite{FGHP??,FGHP13}, which we review in Section~\ref{sec:lz77}: filter the dataset with LZ77, build a standard index for the filtered file, and then create a hybrid of that standard index and an LZ77-based index.  In order to filter the dataset, we assume we are given a reasonable upper bound on the maximum length of any pattern that will be sought.  Our hybrid uses only the simplest component of the LZ77-based index and our preliminary experiments indicated that it is much smaller and also significantly faster than using a standard index by itself.

In Section~\ref{sec:alibi} we describe how to modify our technique to use alignments instead of LZ77.  We call this new index AliBI, for {\it Ali\/}gnment-{\it B\/}ased {\it I\/}index.  It seems reasonable to assume that human genomes will be provided aligned to a reference genomes, and that we will be asked to store their alignments.  Integrating the alignments into the index and using them instead of the LZ77 parse has three main benefits: it saves the space that would be needed to store the alignments separately, while taking advantage of their special structure; it saves us having to compute the LZ77 parse of the whole dataset; and, most importantly, it simplifies and speeds up random access.

\section{LZ77-Based Hybrid} \label{sec:lz77}

Let \(T [1..n]\) be the concatenation of the genomes in the dataset and let $z$ be the number of phrases in the LZ77 parse of $T$.  For our LZ77-based hybrid index we used the variant of LZ77 according to which, for each phrase \(T [i..j]\) in the parse of $T$, either \(i = j\) and \(T [i]\) is the first occurrence of that distinct character, or \(T [i..j]\) occurs in \(T [1..j - 1]\) but \(T [i..j + 1]\) does not occur in \(T [1..j]\).  In the first case, \(T [i]\) is encoded as itself.  In the second case, \(T [i..j]\) is encoded as the pair \((i', j - i + 1)\), where $i'$ is the starting point of the leftmost occurrence of \(T [i..j]\) in $T$; we call this leftmost occurrence \(T [i..j]\)'s source.

Farach and Thorup~\cite{FT95} observed that the first occurrence of any pattern must cross a boundary between phrases.  Occurrences that cross phrase boundaries are called primary and occurrences that are completely contained in a single phrase are called secondary.  In fact, if a phrase contains a secondary occurrence, then the phrase's source must contain an occurrence in the corresponding position.  K\"arkk\"ainen and Ukkonen~\cite{KU96} observed that we can store $T$ in small space such that, once we have found all the primary occurrences of a pattern in $T$, we can quickly find all the occurrences of that substring.

For K\"arkk\"ainen and Ukkonen's approach, we store a data structure for 2-sided range reporting on the \(n \times n\) grid on which there is a marker at \((i, j)\) if \(T [i..j]\) is the source for some phrase; we store a pointer to the starting point of that phrase as satellite data with the marker.  Once we know there is a occurrence (primary or secondary) at \(T [\ell..r]\), we find all phrases whose sources include \(T [\ell..r]\) by finding all markers \((i, j)\) with \(i \leq \ell\) and \(r \leq j\).  If we use $\Oh{z}$ space for the data structure, then a query takes $\Oh{\log \log n}$ time per point reported~\cite{CLP11}.  Therefore, once we have found all the primary occurrences, we can find all $\occ$ occurrences in $\Oh{occ \log \log n}$ time.  This approach has been reused by many subsequent authors, who have thus focused their attention on finding primary occurrences.

\subsection{Primary matches} \label{subsec:primaries}

For our LZ77-based hybrid index, we observed that if we have an upper bound $M$ on the maximum length of any pattern that will be sought and an upper bound $K$ on the maximum edit distance that will be allowed (with $K$ presumably less than $M$) then all primary occurrences of all approximate matches to any pattern will lie entirely within distance \(M + K - 1\) of the nearest phrase boundary.  We discard all characters of $T$ further than \(M + K - 1\) from the nearest phrase boundary, insert \(K + 1\) copies of a special separator symbol \# between each consecutive pair of the resulting substrings, and build a standard index for the resulting string $T_{M, K}$.  Notice that no valid match of any pattern can cross a region of separator characters.  We also build a sorted list saying where each phrase bound in $T$ is mapped to in $T_{M, K}$.

Given a pattern $P$ of length at most $M$ and an edit distance \(k \leq K\), we use the standard index for $T_{M, K}$ to find the primary approximate matches of $P$ in $T$.  To do this, we find all the approximate matches of $P$ in $T_{M, K}$ with the standard index, then use the list mentioned above to discard those approximate matches in $T_{M, K}$ that do not correspond to primary matches in $T$.  More specifically, once we have the start point of an approximate match in $T_{M, K}$, we find via binary search its successor in the list of where the phrase boundaries are mapped to, and check whether the end point of the approximate match is earlier or later.

We then map the remaining approximate matches in $T_{M, K}$ to their corresponding matches in $T$, using a list saying where the phrase boundaries are in $T$.  Notice the offset of the start point of the approximate match in $T_{M, K}$ from the next position where a phrase boundary is mapped to, is the same as the offset of the start point of the corresponding match in $T$ from the next phrase boundary.  For more details, we refer the reader to the preprint of our paper~\cite{FGHP13}.  In summary, finding all the primary approximate matches of $P$ in $T$ takes three data structures:
\begin{itemize}
\item a standard index for $T_{M, K}$,
\item a sorted list of phrase boundaries in $T_{M, K}$,
\item a sorted list of phrase boundaries in $T$.
\end{itemize}
We use gap encoding for both lists, with tables of samples to allow fast access and binary search.

It may be that some of the substrings in $T_{M, K}$ between regions of separator characters are copies of earlier substrings.  We can remove those substrings from $T_{M, K}$ at the cost of adding dummy phrases to the LZ77 parse of $T$.  This is not difficult in theory and seems to improve compression in practice, but complicates the implementation.

\subsection{Secondary matches} \label{subsec:secondaries}

We use K\"arkk\"ainen and Ukkonen's approach of recursive 2-sided range reporting to find secondary matches.  Notice that even if a secondary match of $P$ in $T$ is only an approximate match of $P$, it is an exact copy of an earlier primary match.  Therefore, once we have a list of the primary occurrences of the approximate matches of $P$ in $T$, we can ignore the distinction between exact and approximate pattern matching --- we can find the positions of the secondary matches from the structure of the LZ77 parse without considering $P$ itself anymore.

Rather than using theoretically optimal data structures for 2-sided range reporting, we use a simple and practical solution.  We store a sorted list of the start points of sources, gap-encoded with tables of samples to allow fast access and binary search.  For each source, we store a pointer into the list of phrase boundaries in $T$, mentioned in Subsection~\ref{subsec:primaries}; the phrase corresponding to the source is the one immediately after the indicated boundary.  We do not need to store explicitly the list of the end points of sources, since the length of a phrase's source is the same as the length of the phrase itself, which is the gap from the indicated boundary and the next boundary.  Therefore, as we already have the gap-encoded list of phrase boundaries, we can support access to the list of sources' end points.  However, we store a range-maximum data structure over the list of end points (sorted by start point), which does not use access to the list itself and returns the position in the list of the largest end point in a given range.

Once we have a list of the primary matches, we start a pointer at the head of the list and move it step by step until it reaches the end of the list; whenever we find a secondary occurrence, however, we append it to the tail of the list, so the pointer will pass over it before reaching the end of the list.  When the pointer reaches each match in the list, we perform a 2-sided range reporting query: we using binary search to find the last start point of a source preceding the start point of the match; we then use recursive range-maximum queries and access to the list of end points, to find all the end points of sources whose start points precede the start point of the match.  Again, for more details we refer the reader to the preprint of our paper~\cite{FGHP13}.  In summary, finding all the secondary approximate matches of $P$ in $T$ takes another three data structures:
\begin{itemize}
\item a sorted list of the start points of sources,
\item a list of pointers into the list of phrase boundaries,
\item a range-maximum data structure for the list of sources' end points.
\end{itemize}

Notice it is important that in Subsection~\ref{subsec:primaries} we discard matches in $T_{M, K}$ that do not correspond to primary matches in $T$.  If we do not do this, we will report such matches at least twice, and possibly exponentially many times.

\subsection{Experimental results} \label{subsec:experiments}

We performed preliminary experiments on four files from the repetitive corpus\footnote{http://pizzachili.dcc.uchile.cl/repcorpus.html} on the Pizza\&Chili website.  We chose \(M = 100\) because it seemed like a reasonable pattern length for many bioinformatics applications.  We used a normal FM-index because standard indexes for approximate pattern matching often place restrictions on their input's format or use parallelization, making it difficult to their performance with our preliminary sequential implementation.  Because normal FM-indexes support only exact pattern matching, we set \(K = 0\).

\begin{figure}[t]
\begin{center}
\includegraphics[width=70ex]{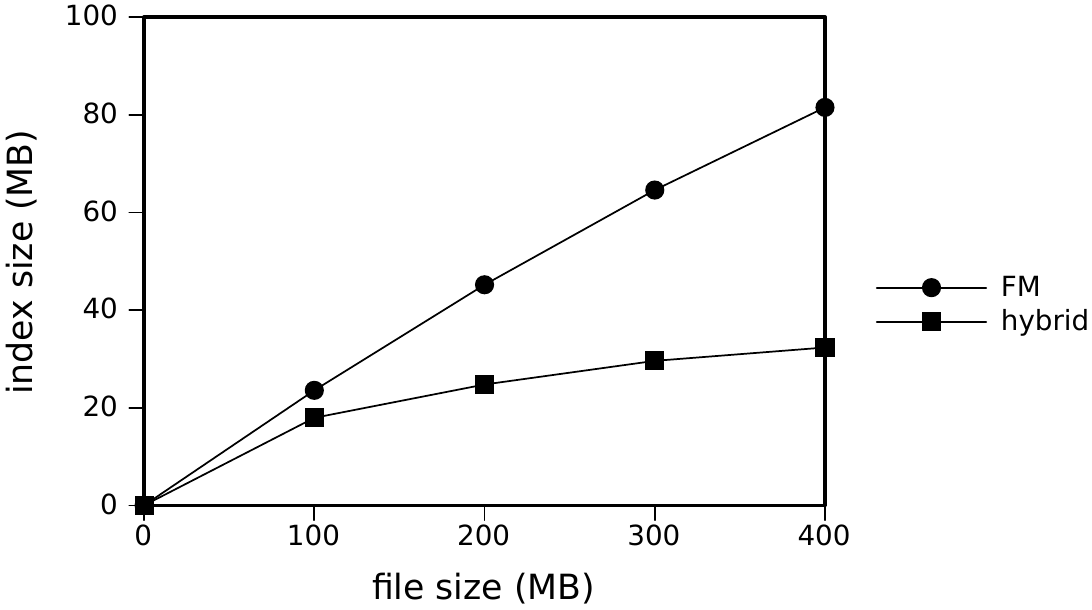}
\end{center}
\caption{Index sizes for prefixes of {\tt cere} of 100, 200, 300 and 400 MB.}
\label{fig:growth}
\end{figure}

For brevity, we discuss here only our results on the {\tt cere} file, which contains the 37 {\em Saccharomyces cerevisiae} genomes from the Saccharomyces Genome Resequencing Project.  The original file is 440 MB, which 7zip compresses to 5 MB; the FM-index\footnote{https://github.com/simongog/sdsl-lite} takes 88 MB; our LZ77-based hybrid index takes 34 MB.  Perhaps more interesting than that, however, is how the FM-index and our index scale, which we tested by storing prefixes of 100, 200, 300 and 400 MB of {\tt cere}.  The results are shown in Figure~\ref{fig:growth}.

\begin{figure}[t]
\begin{center}
\includegraphics[width=70ex]{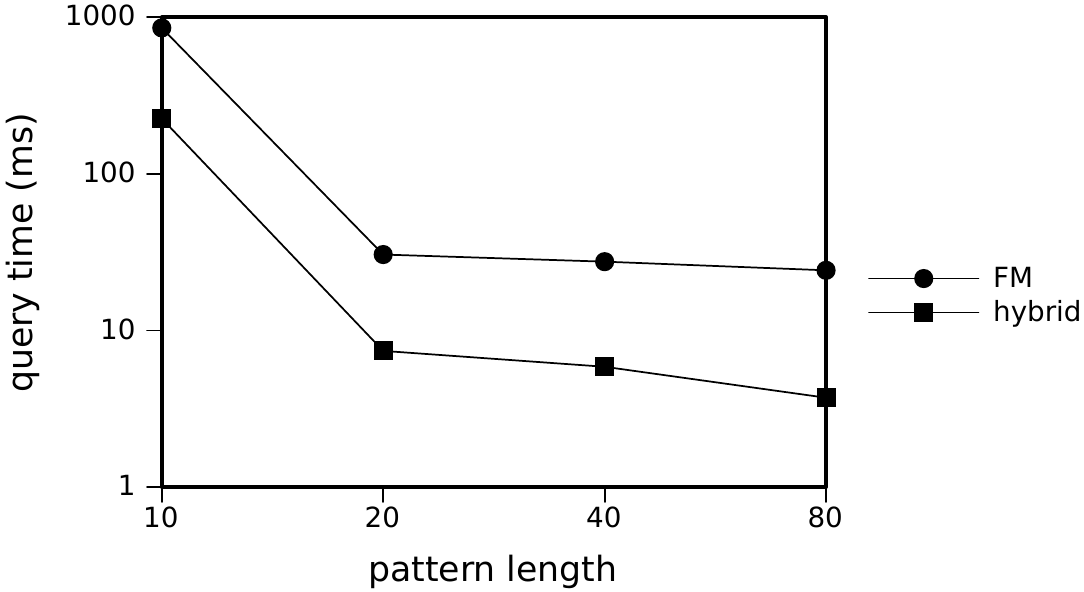}
\end{center}
\caption{Average query times for 3000 patterns.}
\label{fig:queries}
\end{figure}

Of course, it is easy to trade off space for time, but our experiments show that on repetitive datasets, our hybrid is also faster than an FM-index.  Figure~\ref{fig:queries} queries shows the average query time over 3000 patterns selected randomly from the dataset (with unary patterns discarded because they have millions of occurrences) for pattern lengths 10, 20, 40 and 80.  Notice both axes use logarithmic scales.  We believe our index was faster because, for repetitive datasets, most occurrences are secondary, and 2-sided range reporting is faster than the locating operation of FM-indexes.

Figures~\ref{fig:growth} and~\ref{fig:queries} are as they appeared in our previous paper.  Afterwards, we implemented the optimization mentioned at the end of Subsection~\ref{subsec:primaries}, which slightly improved our index's scaling: e.g., for 100 MB our index shrank by 2.5\% but for 400 MB our index shrank by 5\%.

\section{AliBI} \label{sec:alibi}

Two problems with LZ77-based indexing are that computing the LZ77 parse of massive datasets can still be a bottleneck (even with specialized algorithms~\cite{KKP13}) and that LZ77 compression can make random access difficult~\cite{KN13,VY13}.  To overcome these problems, some authors~\cite{DJSS12,GGKNP12} have considered indexing based on a variant of LZ77 called Relative Lempel-Ziv (RLZ)~\cite{KPZ10,KPZ11} which performs well on genomic datasets~\cite{DG11}.  To compute the RLZ parse we choose a reference sequence --- often just the first genome --- and greedily partition the rest of the dataset into phrases that occur in the reference sequence.  This way, while parsing we need store only the reference sequence and the genome currently being parse; for random access, we need only compute in which phrase the desired character lies and its offset in that phrase, then consult the reference sequence.

We considered turning our LZ77-based hybrid index into an RLZ-based hybrid index, but then it occurred to us that human genomes are often aligned to a reference genome anyway before they are stored.  For example, part of an alignment could be written
\[\begin{tabular}{l@{\hspace{3ex}}l}
Reference sequence: & {\tt \dots GATACATTGA---ATCAATCGACGGTT}\,{\it A\/}\,{\tt TGACGGCATATCGCCACATGATA\dots}\\
Aligned sequence: &   {\tt \dots GATACATTGACACATCAATCGACGGTT}\,{\it T\/}\,{\tt TGACGGCATA---CCACATGATA\dots}
\end{tabular}\]
where hyphens in the top line indicate insertions (i.e., characters that appear in the aligned sequence but not the reference sequence), italics indicate substitutions, and hyphens in the bottom line indicate deletions (i.e., characters that appear in the reference sequence but not the aligned sequence).  The differences between human genomes tend to be relatively few, far between and small; single-character substitutions (called single-nucleotide polymorphisms or SNPs) are the most common differences.

For simplicity, we assume we want to index a dataset of human genomes that have been aligned against the first genome. In practice, it should not be difficult to vary this setup by, e.g., excluding the reference genome from the dataset or using several reference genomes.  Given an upper bound $M$ on the maximum length of any pattern that will be sought and an upper bound $K$ on the maximum edit distance that will be allowed, we mark all the characters in each aligned genome within distance \(M + K - 1\) of the nearest character that does not match the corresponding character in the reference genome.  We also mark all the characters in the reference genome.  For example, if \(M = 2\) and \(K = 1\), then in the substring shown above we mark the characters shown below in italics:
\[\mbox{{\tt \dots GATACATT}\,{\it GACACAT\/}\,{\tt CAATCGACGG}\,{\it TTTTG\/}\,{\tt ACGGCA}\,{\it TACC\/}\,{\tt ACATGATA\dots}}\,.\]

Notice that any substring of length at most \(M + K\) that appears in the aligned genome but not the reference genome, consists entirely of marked characters (i.e., is shown entirely in italics).  However, not every substring of length at most \(M + K\) that appears in both genomes consists entirely of unmarked characters (i.e., is shown in plain typeface).  To make this the case, we extend the regions of unmarked characters so they overlapped the marked regions by \(M + K - 1\) characters:
\[\begin{tabular}{cccccccccccc}
& {\it GA} & {\it CAC} & {\it AT} & & {\it TT} & {\it T} & {\it TG} & & {\it TA} & {\it CC} &\\
{\tt \dots GATACATT} & {\tt GA} & & {\tt AT} & {\tt CAATCGACGG} & {\tt TT} & & {\tt TG} & {\tt ACGGCA} & {\tt TA} & {\tt CC} & {\tt ACATGATA\dots}\,.
\end{tabular}\]

In this context, we call an occurrence of a substring secondary if it lies entirely in a region of unmarked characters; otherwise, we call it primary.  In the example above, there are three secondary occurrences of {\tt TG} but no primary occurrences of it --- the occurrence at the end of the second substring in italics is in the overlap between marked and unmarked characters, so it is still secondary and not primary --- while there are three primary occurrences of {\tt AC} and four secondary occurrences of it.

We concatenate all the marked substrings, including the reference genome, with each pair separated by \(K + 1\) copies of \#.  To save space, we include each distinct marked substring only once; for each distinct marked substring, we store a list of pointers to where it occurs in aligned genomes.  A marked substring is likely to occur in roughly the same place in all the genomes in which it appears, so we can compress these pointers; we will include more details in the full version of this paper.

We build a standard index for this string concatenation, which we assume supports approximate pattern matching as well as fast random access; otherwise, we can store a separate data structure for random access.  With this index we can quickly find all the primary approximate matches of a given pattern $P$ in $T$.  To do this, search the index and discard any match that is completely contained in the margins of length \(M + K - 1\) of a marked substring, since that match is actually secondary.

To be able to find $P$'s secondary approximate matches from the list of the primary matches in the reference genome, we again store a data structure for 2-sided range reporting on a grid.  This time, however, the size of the grid is bounded in both directions by the size of the reference genome, rather than the size of the whole dataset.  Moreover, since an unmarked substring is likely to occur in roughly the same place in the aligned genome as in the reference genome, we can also compress our range reporting data structure.  Again, we will give more details in the full version of this paper.

We hope to have a prototype of this alignment-based index, AliBI, implemented and tested by early in 2014.  We expect it to save space by integrating the alignments into the index and using them instead of the LZ77 parse because we need not store the indexes separately and we can take advantage of their special structure.  We also expect the time to construct the index to be reduced, since we need not compute the LZ77 parse of the whole dataset, and we expect random access to be faster.

\section*{Acknowledgments}

Many thanks to Pawe\l\ Gawrychowski, Juha K\"arkk\"ainen, Veli M\"akinen, Gonzalo Navarro and Jorma Tarhio, for helpful discussions.  This research was funded by Fondecyt grant 1-110066, the Helsinki Institute for Information Technology, and Academy of Finland grants 134287 and 258308.

\end{document}